\documentclass[twocolumn,aps,pre,superscriptaddress,longbibliography]{revtex4-2}
\usepackage[utf8]{inputenc}
\usepackage[english]{babel}
\usepackage{amssymb, amsmath, braket, hyperref, graphicx, color}
\usepackage{natbib}

\begin{document}
\title{Exploring Quantum Annealing Architectures: A Spin Glass Perspective}

\author{Gabriel Jaumà}
\affiliation{Institute of Fundamental Physics IFF-CSIC, Calle Serrano 113b, 28006 Madrid, Spain}
\affiliation{Applied Physics Department, Salamanca University, Salamanca 37008, Spain}

\author{Juan José García-Ripoll}
\affiliation{Institute of Fundamental Physics IFF-CSIC, Calle Serrano 113b, 28006 Madrid, Spain}

\author{Manuel Pino}

\affiliation{Nanotechnology Group, USAL-Nanolab, Salamanca University, Salamanca 37008, Spain}
\affiliation{Institute of Fundamental Physics IFF-CSIC, Calle Serrano 113b, 28006 Madrid, Spain}

\begin{abstract}
We study the spin-glass transition in several Ising models of relevance for quantum annealers. We extract the spin-glass critical temperature by extrapolating the pseudo-critical properties obtained with Replica-Exchange Monte-Carlo for finite-size systems. We find a spin-glass phase for some random lattices (random-regular and small-world graphs) in good agreement with previous results. However, our results for the quasi-two-dimensional graphs implemented in the D-Wave annealers (Chimera, Zephyr, and Pegasus) indicate only a zero-temperature spin-glass state, as their pseudo-critical temperature drifts towards smaller values. This implies that the asymptotic runtime to find the low-energy configuration of those graphs is likely to be polynomial in system size, nevertheless, this scaling may only be reached for very large system sizes---much larger than existing annealers---, as we observe an abrupt increase in the computational cost of the simulations around the pseudo-critical temperatures. Thus, two-dimensional systems with local crossings can display enough complexity to make unfeasible the search with classical methods of low-energy configurations.
\end{abstract}
\maketitle

\section{Introduction}

Quantum annealing is based on the minimization of a classical cost function using quantum fluctuations\ \cite{apolloni_quantum_1989,kadowaki_quantum_1998,albash_adiabatic_2018, hauke_perspectives_2020}. This algorithm can be implemented in hardware constructed with several quantum technologies as superconducting circuits\ \cite{nakamura_coherent_1999,johnson_quantum_2011, maezawa_toward_2019}, Rydberg atoms\ \cite{glaetzle_coherent_2017,serret_solving_2020}, optical lattices\ \cite{torggler_quantum_2017, pino_capturing_2013} or trapped ions\ \cite{nevado_hidden_2016,gras_quantum_2016}, having each of them their own strengths and limitations. Among these platforms, superconducting circuits have shown one of the best performances, with processors reaching up to thousands of qubits \cite{teplukhin_computing_2021}, and coherent dynamic in subsystems comprising as many as $20$ sites\ \cite{king_coherent_2022}. 

\begin{figure*}[t!]
    \centering
    \includegraphics[width=0.9\textwidth]{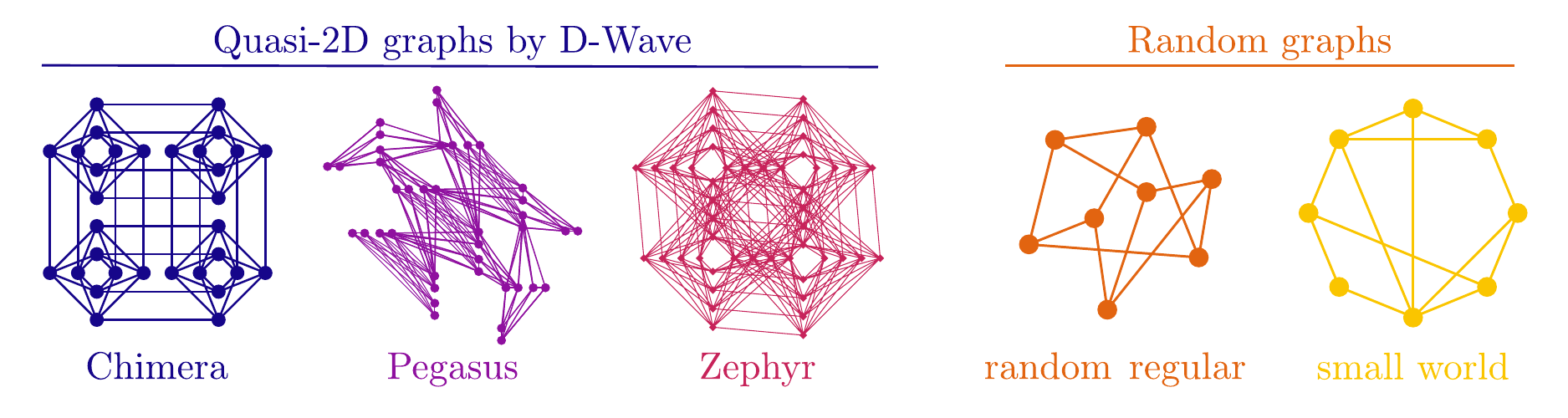}
    \caption{Schematic representation of the different graph types analyzed in this study. This includes the three D-Wave topologies: Chimera, Pegasus, and Zephyr, along with two random graphs: $k$-random regular ($k$-RRG) and small world graphs.}
    \label{fig:architectures}
\end{figure*}

Quantum annealers for stoquastic problems face relevant challenges to demonstrate a quantum advantage over classical optimization algorithms. At the hardware level, adiabatic passages must be performed with an adequate speed to guarantee a large overlap with the desired ground state\ \cite{boixo_experimental_2013,king_coherent_2022}. Still, the ground state itself may experience chaos due to small changes in the problem Hamiltonian\ \cite{martin-mayor_unraveling_2015}, a similar phenomenon than \textit{temperature chaos} in spin glasses\ \cite{lukic_temperature_2006,fernandez_temperature_2016}. At the software level, the challenge is to identify families of problems and concrete instances that can be solved more efficiently in the quantum annealer than in classical solvers and to demonstrate, if not better scaling, significantly better prefactors in the resources\ \cite{boixo_experimental_2013,boixo_evidence_2014,heim_quantum_2015,albash_demonstration_2018,king_coherent_2022}.

In this work, we adopt an alternative point of view that unifies the hardware and software limitations from a statistical physics point of view\ \cite{katzgraber_monte_2001,katzgraber_seeking_2015,liers_ground_2003,liu_quantum_2015}. This involves analyzing the  Ising models that can be most naturally implemented in the annealer, studying their phase diagram, and the classical hardness of extracting low-energy properties for different topologies or connectivities. Using parallel tempering, we compute the pseudo-critical temperature---or crossover temperature at finite systems---for spin glasses on quantum annealer topologies with $k$-random regular graphs ($k$-RRG), small-world networks, and the quasi-2D topologies of D-Wave annealers [cf. Fig.\ \ref{fig:architectures}]. In quasi-2D D-Wave topologies, our simulations are compatible with a spin-glass phase restricted to only $T=0,$ as the pseudo-critical temperature seems to flow to zero in the thermodynamic limit. At the same time, the pseudo-critical temperature in $k$-RRG and small-world networks converges to a non-zero critical value, giving rise to a finite spin-glass phase.

Our previous findings may suggest that quasi-2D topologies are computationally "trivial" and small-world architectures are more advantageous for quantum annealing\ \cite{katzgraber_how_2018}, as Monte-Carlo based methods are only expected to have trouble in the spin-glass phase of our stoquastic models. However, the analysis of the autocorrelation times in parallel tempering reveals a divergent scaling of the running time of parallel tempering, suggesting that classical optimization could be challenged by a quantum annealer in practical situations and finite-size problems. This is consistent with previous results about the nature of the 2D-spin-glass phase\ \cite{bray_lower_1984,hartmann_ground_2011,fernandez_universal_2016,fernandez_experiment-oriented_2019} because the pseudo-critical temperature decreases slowly with the problem size for quasi-2D lattices. Furthermore, our analysis can be used to gauge the relative complexity of different D-wave architectures, revealing the enhanced difficulty of more recent ones (Zephyr, Pegasus) as compared to previous annealers (Chimera).

The outline of this work is as follows. In Sect.~\ref{sec:model} we introduce the models under study, and the classification of the topologies under study. We also present the method and quantities that we use to characterize the phase transitions. We analyze the phase diagram of the previous models in Sect.\ \ref{sec:Phase_diagram}. This section is divided into two parts. In the first one~\ref{sec:finite-size-scaling}, we introduce the notion of pseudo-critical temperature as a proxy value that identifies the change of phases in finite-size systems and briefly discuss our technique to extract the critical temperature.  In the second \ref{sec:pseudo-critical}, we use the scaling of the pseudo-critical temperature to study the phase transitions for five families of graphs: $k$-random regular graphs, small world networks, usually regarded as mean-field problems, and D-Wave's Chimera, Pegasus and Zephyr architectures [cf. Fig\ \ref{fig:architectures}]. We then analyze the time needed to perform parallel tempering via the autocorrelation in Sec.\ \ref{sec:auto}. In Sec.\ \ref{sec:concl}, we summarize our results, emphasizing how these techniques contribute to the search for classically hard-to-solve architectures for quantum annealing.

\section{Model and methods}
\label{sec:model}

We study the spin-glass Hamiltonian
\begin{equation}
    \mathcal{H} = \sum_{ \braket{ij} } J_{ij} S_i S_j,
    \label{eq:model}
\end{equation}
where $S_i\in\set{\pm1}$ are $N$ classical 1/2-spins and the magnitude of the couplings $J_{ij}$ are random variables chosen from a Gaussian distribution with zero mean and unit variance. The sum runs over all pair of spin indexes that are coupled following a given graph $G.$ That is, each spin and each coupling represents a node and an edge in the graph, respectively. 

We study some random graphs with an effective infinite dimension such as $k$-RRG\ \cite{liers_ground_2003} and small-world networks \cite{katzgraber_glassy_2014}. Both of them look locally as the branching of a tree but differ from a pure tree (Bethe lattice) in the presence of loops of length that scales as $\log(N)$\ \cite{dorogovtsev_critical_2008}, see Fig.\ \ref{fig:architectures}. The connections for a $k$-RRG are randomly chosen but with the restriction that each spin should be coupled to exactly $k+1$ other spins, $k$ being the branching number. The small world graphs are constructed by connecting all the spins in a ring, and then adding connections randomly until the average number of neighbors per spin is $k+1$.

We will also compute Ising Hamiltonians on the graphs used by the D-Wave processors. Those are the three quasi-2D graphs referred to as Chimera, Pegasus, and Zephyr (see Fig.\ \ref{fig:architectures}). The difference between them lies in the number and length of the connections. For instance, the older version of the D-Wave processors used the Chimera graph which can be constructed as blocks of eight spins, which are strongly connected between them, but the connections between blocks follow a rectangular lattice, see Fig.\ \ref{fig:architectures}. Further generations of D-Wave processors employ Pegasus and Zehpyr graphs, in which the range of connections is increased significantly, although they still flow under the renormalization group to a 2D structure. 

We characterize the phase transition of the Ising models in the previous graphs via the Binder cumulant\ \cite{binder_spin_1986}.  This quantity is defined as the fourth order cumulant of the probability distribution of the order parameter $q$\ \cite{iniguez_3d_1997}:
\begin{equation}
    g = \frac{1}{2}\left(3-\frac{\overline{\braket{q^4}}}{\overline{\braket{q^2}}^2} \right),
    \label{e. def Binder cumulant}
\end{equation}
where $q$ is the overlap of two spin configurations, $S_i^\alpha$ and $S_i^\beta$, resulting from two independent simulations of the same Hamiltonian at a given temperature:
\begin{equation}
    q = \frac{1}{N}\sum_i S_i^\alpha S_i^\beta\;.
\label{eq:spin-overlap}
\end{equation}
The bracket in Eq. \ref{e. def Binder cumulant}  indicates a thermal average, and the overline indicates an average over realizations of the couplings $J_{ij}$ in Hamiltonian Eq. \ref{eq:model}. 

\begin{figure*}[t!]
    \centering
    \includegraphics[width=.95\textwidth]{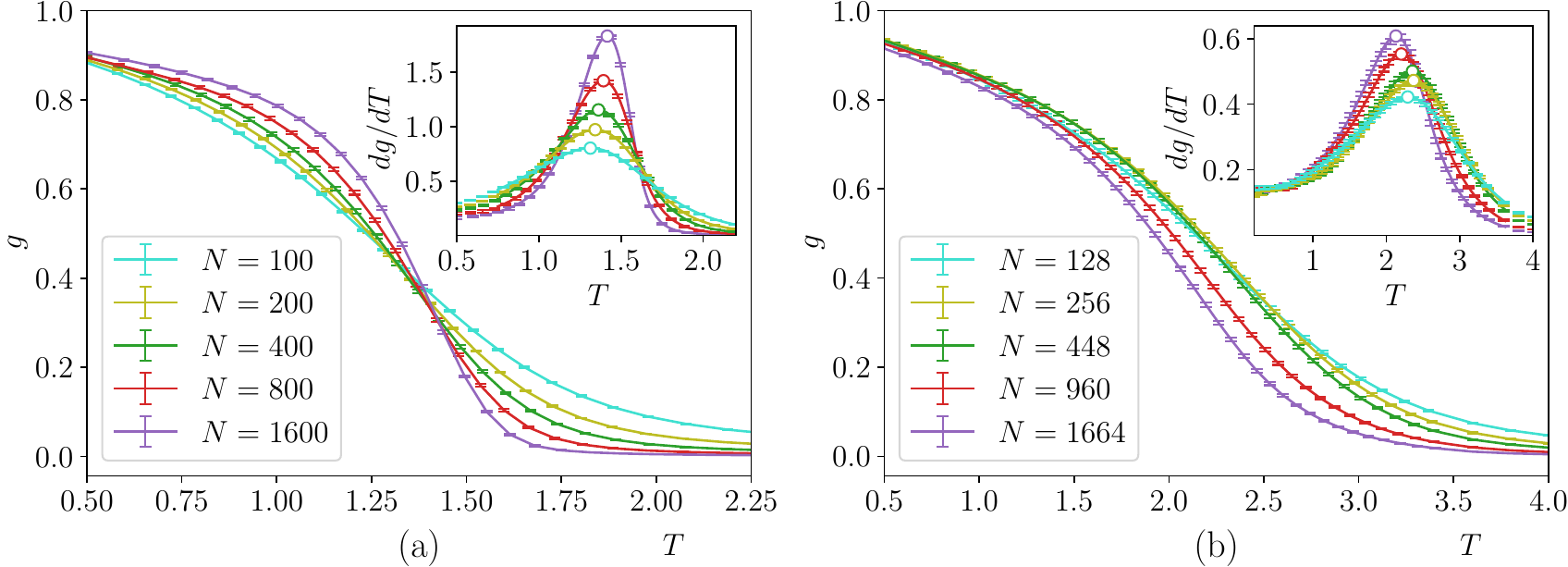}
    \caption{Binder cumulant $g$ as a function of temperature for \textbf{(a)} $4$-RRG and \textbf{(b)} Pegasus for several system sizes. The solid lines are guides to the eyes. The crossing of $g$ for different sizes at the $4$-RRG suggests a phase transition around $T_C\sim1.4$. Conversely, the absence of such crossing at Pegasus hints at a zero critical temperature. Inset: numerical derivative of the Binder cumulant together with their Padé approximants (solid lines). The peaks (marked with circles) indicate the pseudo-critical temperature for each size.}
    \label{fig:FSS-g-4RRG}
\end{figure*}

We use the Markov chain Monte Carlo method (MCMC)\ \cite{metropolis_equation_2004} to calculate thermal averages by transforming them into temporal averages taken from a Markov chain in thermal equilibrium. We use the parallel tempering algorithm\ \cite{hukushima_exchange_1996}, which simultaneously simulates $N_T$ Markov chains for each realization at a different temperature and allows them to swap temperatures. These swaps help to reduce thermalization times by allowing low-temperature Markov dynamics to escape local minima. We have simulated around 10000 realizations of each graph for system sizes up to 1600 sites and as long as 10 million Monte Carlo Sweeps (MCS). We have used the bootstrap method \ \cite{efron_bootstrap_1992} to compute error bars in several quantities. This technique allows us to deal at the same time with errors coming from thermal and different realizations averages. 

% We do so because error-propagating techniques result in very convoluted expressions: on the one hand, our numerical simulations have two sources of error, the statistical average errors and the thermal average errors; on the other hand, quantities such as the binder cumulant involve non-linear combinations of statistical and thermal averages of correlated quantities.

\section{Phase diagram}
\label{sec:Phase_diagram}

We have performed a finite-size scaling analysis of the  Binder-cumulant's derivative for the graphs under study. We have found a phase diagram with a finite spin-glass phase for the random-regular and small-world graphs---in agreement with the results from a mean-field analysis \cite{thouless_spin-glass_1986}---while our results indicate a spin-glass phase restricted to just the $T=0$ point. Our main tools to do so have been the characterization of the pseudo-critical points\ \ref{sec:finite-size-scaling} and the extrapolation of those points to the thermodynamical limit\ \ref{sec:pseudo-critical}

\begin{figure*}[t!]
    \centering
    \includegraphics[width=0.9\linewidth]{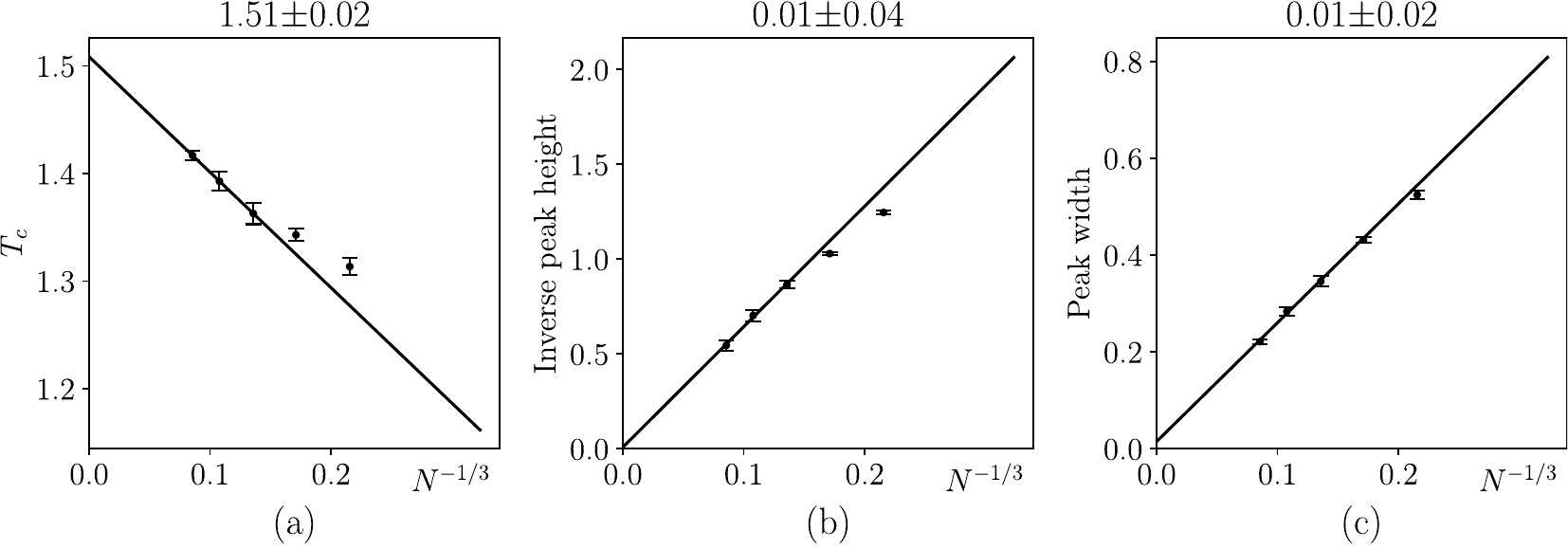}
    \caption{Finite-size scaling analysis of the properties of the derivative of the Binder cumulant for $4$-RRG as a function of the system size. (a) Position of the peak (pseudo-critical temperature), (b) inverse peak height, and, (c) peak width, calculated as the width at $80\%$ height. The solid lines are linear fits for the three largest sizes, and the titles correspond to the intersection of the line with the $y$ axis, which is the value of each quantity in the thermodynamic limit.}
    \label{fig:FSS-Tc-4RRG}
\end{figure*}

\subsection{Pseudo-critical temperatures at finite systems}
\label{sec:finite-size-scaling}

We explain how to characterize the pseudo-critical temperature for finite system sizes. That pseudo-critical temperature $T_{C}(N)$ is the one at which a finite system of size $N$ crossover from a paramagnetic (high temperature) to a spin-glass regime (low temperature). To find it, we use the Binder cumulant introduced in the previous section, which changes from $0$ to a finite value when going from the paramagnet to the spin-glass regime. Although this change is smooth for a finite system, it can become steeper and steeper upon increasing size, giving rise to a discontinuity at infinite system sizes. We will extract the pseudo-critical temperature as the one where the maximum in the Binder cumulant occurs at a given size\ \cite{cardy_scaling_1996}.  As we will see, the location of the true critical point at an infinite system size can be obtained by performing a finite-size scaling of those pseudo-critical temperatures. 

Specifically, the maxima in the Binder cumulant are extracted by estimating the derivative using second-order central differences and then fitting the result using Padé approximants with a degree that minimizes the quantity $|1-\chi^2|,$ similar to Ref.\ \cite{pino_scaling_2020}. The position of this maxima signals the pseudo-critical points for each size. To exemplify this, the Binder cumulant for a $4$-RRG and Pegaus graphs are represented in Fig.\ \ref{fig:FSS-g-4RRG}. In the inset of each panel, we have also presented Binder's cumulant derivative, highlighting with a circle the position of their maxima. The position at which this maximum occurs marks the location of the pseudo-critical temperature.

There is an important difference between the data for 4-RRG and Pegasus, Fig.\ \ref{fig:FSS-g-4RRG}. For the first case, there is a well-defined crossing point of the Binder cumulant for different sizes, while this crossing point is absent in the data for Pegasus.  In the case of a stable crossing point, it is relatively easy to extract critical properties by collapsing the Binder cumulant to a scaling form $A\sim \xi^{\zeta} f(N^{1/\nu} (T-T_c))$ being $f$ a scaling function\ \cite{rodriguez_critical_2010,pino_ergodic_2019,katzgraber_how_2018, cardy_scaling_1996}. However, this method is more cumbersome\ \cite{katzgraber_how_2018} for the quasi-2D graphs due to the absence of this crossing point. For that reason, we have based all our analysis on the determination of the pseudo-critical temperature. We will see that all relevant critical properties can be determined by analyzing this quantity's scaling.

\subsection{Critical temperatures}
\label{sec:pseudo-critical}

\subsubsection{Random-regular and small-world graphs}

We explain our procedure to extract the critical temperature for the case of random-regular graphs with branching number $k=4$, as the analysis of the other graphs is the same. The peak position (pseudo-critical temperature), peak height, and peak width of the derivative of the Binder cumulant are shown in panels (a), (b), and (c) of Fig.\ \ref{fig:FSS-Tc-4RRG}. A scaling exponent $\nu=3$ is used to extrapolate those properties to the thermodynamic limit, see Fig.\ \ref{fig:FSS-Tc-4RRG}. Our results show that a discontinuity of the Binder cumulant derivative occurs for $N\rightarrow\ \infty$ at the critical temperature, which is the extrapolated value of the pseudo-critical ones). The exponent $\nu=3$ comes from the fact that $\nu=\nu_{\rm MF}d_{\rm u}$, where $d_{\rm u}=6$ is the upper critical dimension of the Ising spin-glass transition \ \cite{liers_ground_2003,katzgraber_how_2018} and $\nu_{\rm MF}=1/2$ is the mean-field exponent \ \cite{fischer_spin_1991}. Looking at Fig. \ref{fig:FSS-Tc-4RRG} we can see that the scaling function contains additional finite size corrections to the law $1/N^{-1/3}$, and to minimize their effects, we have performed our analysis only with the three largest sizes. 

\begin{figure}[t!]
    \centering
    \includegraphics[width=.455\textwidth]{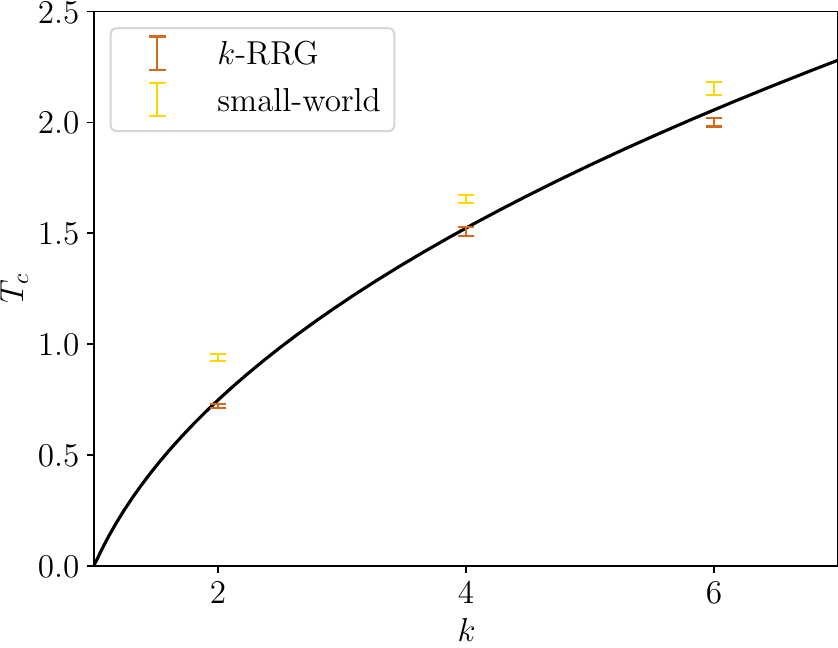}
    \caption{Critical temperatures as a function of the branching number $k$ for RRGs and small-world networks. The solid line is the theoretical prediction from Ref.\ \cite{mezard_bethe_2001}. Note that the theoretical prediction is strictly valid for graphs with a constant branching number, explaining the discrepancy with small-world networks where the branching number is a random variable.  }
    \label{fig:Tc-numeric-vs-theory-RRG-SW}
\end{figure}

We have repeated this procedure for small-world graphs and found that they have the same qualitative behavior: a critical exponent $\nu=1/3$ that controls the divergence of the Binder cumulant. The extrapolated critical temperatures for all the random graphs have been plotted in Fig.\ \ref{fig:Tc-numeric-vs-theory-RRG-SW} as a function of the branching number together with the theoretical prediction from Ref.\ \cite{mezard_bethe_2001}. This theoretical prediction is obtained as the temperatures where a non-zero average value appears for the distribution probability, the one that is self-consistently found, of local fields.

Analytical results agree for 2-RRG and 4-RRG but differs slightly for 6-RRG. We believe that this is because the additional finite size corrections to the law $1/N^{-1/3}$ become more relevant upon increasing branching number $k$. Additionally, the analytical prediction differs from the numerical results for small-world networks. This is because it only applies to graphs with a constant branching number while, in small-world networks, the branching number is a variable that can take on different values from node to node. A more precise way of characterizing analytically the critical temperature in those networks would be to use a population dynamic algorithms\ \cite{mezard_bethe_2001}. 
 
\subsubsection{D-Wave's graphs}

We have realized a similar analysis for the D-wave graphs as we did for the random ones. Two characteristics of the D-wave graphs depart from the behavior seen in random-regular and small-world graphs [cf. Fig.\ \ref{fig:FSS-g-4RRG}]. We have already commented that there is a lack of a crossing point for the Binder cumulant for different sizes, and second, the tendency of the pseudo-critical temperature is non-monotonous, that is, the slope $\frac{dT_c}{dN}$ changes from being positive to negative upon increasing system size. The lack of a crossing point and the change of tendency of the pseudo-critical temperature appears for all of the D-Wave graphs. They are compatible with a zero critical temperature or, in other words, with a spin-glass state only at zero temperature. However, we have not been able to obtain reliable extrapolations for the critical temperature at the thermodynamical limit due to limitations in system sizes.

To clarify all we said in the previous paragraph, we have plotted the pseudo-critical temperatures for D-wave graphs in Fig\ \ref{Fig:Tc-DWave}. We can see the non-monotonic drift of those temperatures upon increasing system size. This behavior is more clearly seen in the Zephyr graph, which is the most complex one in terms of the range and number of connections. Another important observation is that the pseudo-critical temperature increases when going from Chimera to Zephyr graphs. The graph with the higher pseudo-critical temperature is more beneficial for performing quantum annealing, as the noise on the devices implementing this algorithm can make the exploration of a spin-glass regime at too low a temperature impracticable.

To understand our results regarding the D-wave graphs, we notice that they all show a common feature: they can be laid in a plane with crossing edges only at a local scale. That is, by zooming out sufficiently and renormalizing the spin variables, one would obtain a two-dimensional graph with only local couplings. In terms of the renormalization group, this means that the critical properties defined by the long-distance behavior are those of 2D systems. Indeed, our numerical results indicate that Ising models in D-Wave graphs share key properties with the ones of the Ising spin-glass in two dimensions, as having a zero critical temperature\ \cite{bray_lower_1984,hartmann_ground_2011, fernandez_universal_2016, fernandez_experiment-oriented_2019, franz_interfaces_1994, carmona_critical_1998, demirtas_lower-critical_2015, maiorano_support_2018}.

\section{Autocorrelation times in the Markov chain dynamics}
\label{sec:auto}

Up to now, we have obtained results that may indicate that the graphs used by D-Wave could be easy to anneal via classical methods. If the pseudo-critical temperature decreases towards zero, the spin-glass state only exits at zero temperature. Monte Carlo may then be used in the paramagnetic phase $T>0$ to obtain arbitrarily low-energy configurations. However, this need to be reconciled with the difficulties we have faced to perform the replica exchange Monte-Carlo for low temperatures and sizes $N>3000$ for any of the D-Wave graphs. 

\begin{figure}[t!]
    \centering
    \includegraphics[width=.43\textwidth]{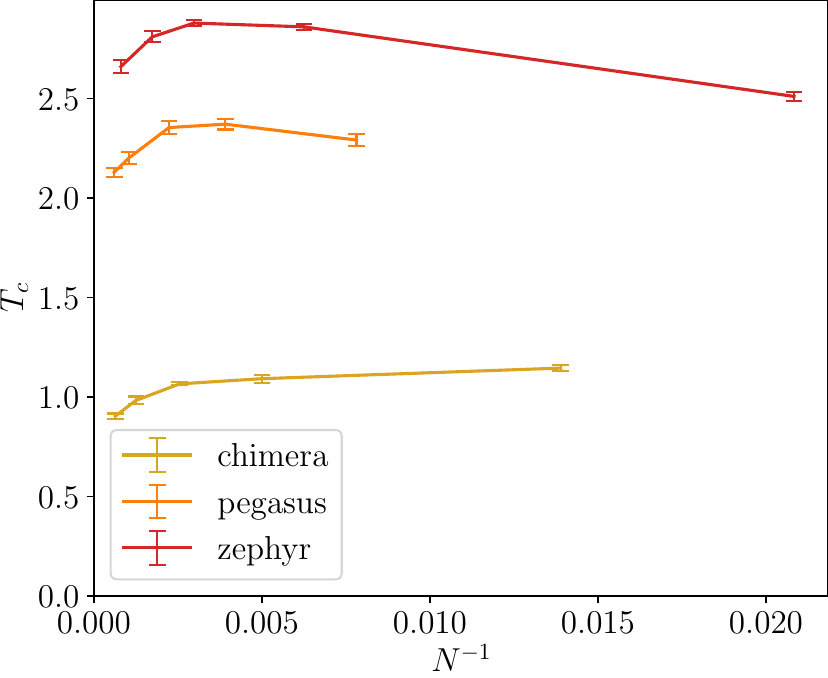}
    \caption{Pseudo-critical temperature as a function of the inverse of system size for the D-Wave's Chimera, Pegasus, and Zephyr graphs. Solid lines are visual guides.
    }
    \label{Fig:Tc-DWave}
\end{figure}

We characterize the difficulties of running parallel tempering in our models by looking at the autocorrelation time of the square of the spin overlap $\tau_{q^2}$ along one Markov chain, see Appendix \ref{appendix:correlation}). This will allow us to correlate the appearance of the pseudo-critical temperature with a substantial increase in the time required to thermalize our parallel tempering simulations. We will also obtain differences in the growth of these times between the graphs studied here.

\begin{figure*}[t!]
    \centering\includegraphics[width=\textwidth]{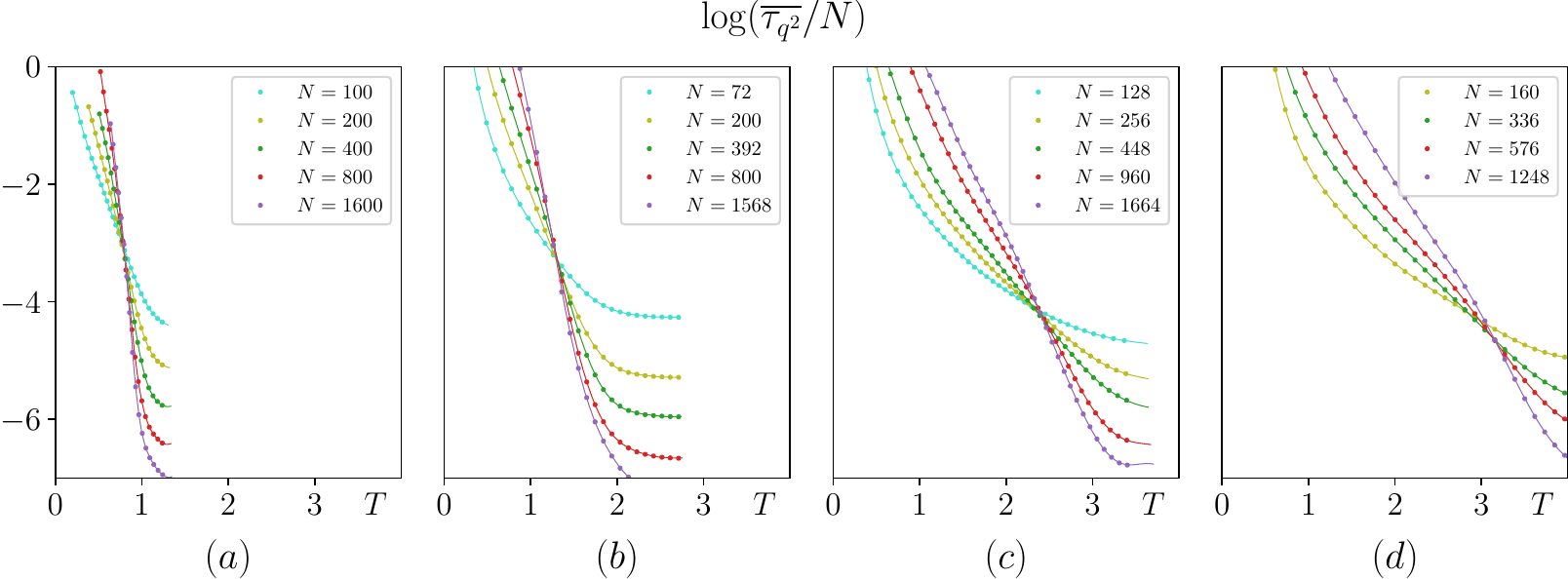}
    \caption{Autocorrelation times $\tau_{q^2}$ as a function of temperature for different sizes and graph architectures (a) 2-RRG, (b) Chimera, (c) Pegasus, and (d) Zephyr.}
    \label{fig:autocorrelation-times-2RRG-and-DWave}
\end{figure*}

In Fig.\ \ref{fig:autocorrelation-times-2RRG-and-DWave}, we have plotted the autocorrelation times $\tau_{q^2}$ divided by system size $N$ obtained from our Markov chain Monte-Carlo for RRG (a), Chimera (b), Pegasus (c) and Zephyr (d) graphs. In all of those plots, a crossing point indicates that the system crossover from a sub- to a super-linear law. The temperature where this occurs is correlated with the pseudo-critical point found earlier, see Figs. \ref{fig:Tc-numeric-vs-theory-RRG-SW} and \ref{Fig:Tc-DWave} . Roughly speaking, all the correlated times begin to increase at temperatures that are slightly above the pseudo-critical points found earlier.

We have tried to characterize further the increase of autocorrelation times in Fig.\ \ref{fig:autocorrelation-times-2RRG-and-DWave} by fitting them to a function of size as $\overline{\tau}_{q^2}= C(T)\left(N/N_0(T)\right)^{A(T)}$ for each temperature. We see a steep increase in each graph's exponent $A(T)$. Furthermore, this growth happens near the pseudo-critical points (star) found earlier in Figs. \ref{fig:Tc-numeric-vs-theory-RRG-SW} and \ref{Fig:Tc-DWave}. It is also relevant to note that RRG presents the steepest increase of $A(T)$ upon approaching the transition, while Zephyr exhibits the less pronounced one. We have already seen that the pseudo-critical point does not change significantly with size for any of the graphs treated. Hence, the differences in the slopes may be related to the structure of the energy landscape in the configuration space. The results in figures\ \ref{fig:autocorrelation-times-2RRG-and-DWave} and \ref{fig:autocorrelation-time-power-law} contain strong indications that Monte-Carlo simulations begin to find difficulties around the pseudo-critical points found earlier, not around the true critical point in the thermodynamic limit. 

We have tried to see if our data for correlations times $\tau_{q^2}$ could be described with an exponential law instead the simpler law used before; see Appendix\ \ref{appendix:correlation}. One may expect this to be the case, at least for RRG, where there is a spin-glass phase at low temperatures\ \cite{mezard_bethe_2001, liers_ground_2003}, as spin-glasses are NP-hard problems\ \cite{barahona_computational_1982}.  Our data seems to contain a curvature when plotting $\log(\tau_{q^2})$ as a function of $\log(N)$, which is compatible with an exponential law below the transitions for all the models studied, see Appendix\ \ref{appendix:correlation}. However, we have found problems when fitting to such a law due to limitations in system size, which makes it difficult to assess the goodness of fit and compare it with the polynomial law employed in Fig.\ \ref{fig:autocorrelation-times-2RRG-and-DWave}. 

We notice that the characterization of the difficulties of Ising problems in our graphs can be subtle, as one may argue that it depends on the method employed. However, our procedure can be justified taking into account that Monte-Carlo methods have been the direct competitor of quantum annealers in many cases\ \cite{martin-mayor_unraveling_2015,katzgraber_glassy_2014} and it is one of the best general methods for solving QUBO problems. 

\section{Conclusions}\label{sec:concl}

Our main conclusion is that quasi-2D architectures, such as those employed by D-Wave, are a good playground to explore the potential of quantum annealers to exhibit a quantum advantage. A crucial piece of information supporting this conclusion is the slow decrease of pseudo-critical temperature with problem size. This behavior causes problems for classical Markov Chain Monte Carlo methods, which experience difficulties exploring lower-energy configurations. Indeed, we have analyzed autocorrelation times for parallel tempering---a rough estimation of the simulation run time---observing a steep increase around the position of the pseudo-critical temperature.

This conclusion may seem at odds with previous works\ \cite{katzgraber_glassy_2014, katzgraber_seeking_2015}. Even if finding the ground state of a quasi-2D system is an NP-hard problem\ \cite{barahona_computational_1982}, the time of finding low-energy configuration via classical methods can scale polynomially for large enough system sizes. However, our results suggest that may not be the case for finite system sizes, as Monte Carlo also experiences difficulties below and around its pseudo-critical temperature. In any case, it could be the case that quantum annealing on quasi-2D graphs cannot beat parallel tempering, but if so, it is likely due to other limitations. For instance, too short adiabatic passages\ \cite{king_coherent_2022}, chaos on the ground-state configuration\ \cite{martin-mayor_unraveling_2015}, a worse performance of quantum tunneling in comparison to thermal fluctuations, or simply the fact that certain stoquastic adiabatic passages can be simulated classically\ \cite{ciani_stoquasticity_2021,halverson_efficient_2020}.

\begin{figure}[t!]
    \centering\includegraphics[width=.45\textwidth]{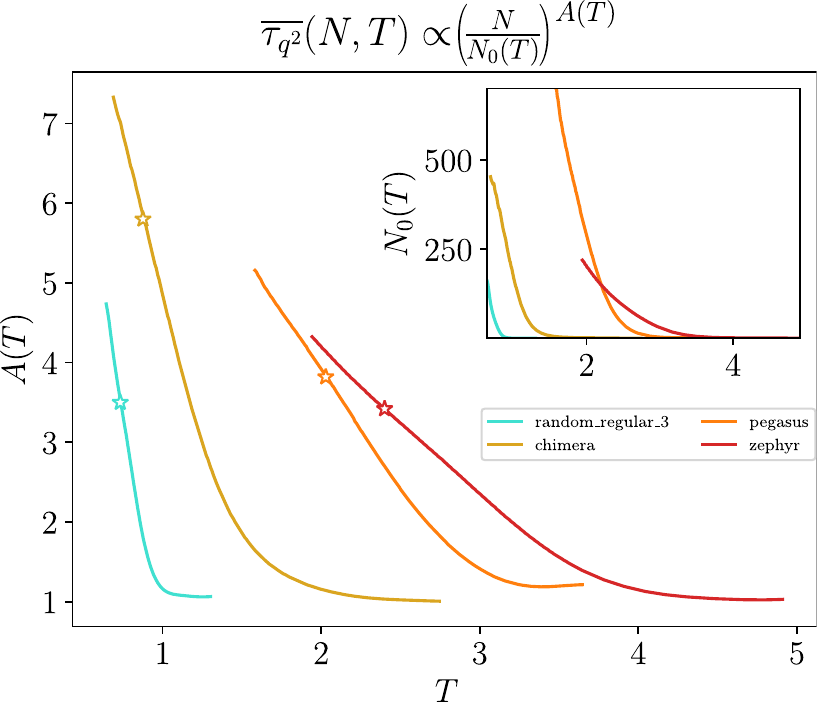}
    \caption{Effective power of the autocorrelation time as a function of temperature. The effective power is extracted from a fitting of the autocorrelation times as a function of system size $\overline{\tau}_{q^2}= C(T)\left(N/N_0(T)\right)^{A(T)}$ for each temperature.}
    \label{fig:autocorrelation-time-power-law}
\end{figure}

As an outlook for future exploration, it would be interesting to continue this study to understand which quasi-2D graphs and topologies produce the most challenging problems. A proxy measure for such a study could be searching the architectures that produce the largest pseudo-critical temperature. It would also be interesting to explore which topologies lead to the steepest increase in autocorrelation times around the pseudo-critical temperature and understand whether the behaviors we have observed [cf. Fig.\ \ref{fig:autocorrelation-time-power-law}] can be associated with some type of universal behavior.

\begin{acknowledgments}
The project that gave rise to these results received the support of a fellowship from “la Caixa” Foundation (ID 100010434). The fellowship code is “LCF/BQ/DR22/11950032”. This work has also been supported by European Commission FET-Open project AVaQus GA 899561, CSIC Quantum Technologies Platform PTI-001. M. P. acknowledges support by Spanish MCIN/AEI/10.13039/501100011033 through Grant No. PID2020-114830GB-I0. The numerical computations have been performed in the cluster DRAGO of the CSIC, the clusters FinisTerrae II and FinisTerrae III from the Galician Supercomputing Center (CESGA), and the facilities of the Supercomputación Castilla y León (SCAYLE), which have been funded by the Spanish Ministry of Science and Innovation, the Galician Government and the European Regional Development Fund (ERDF).
\end{acknowledgments}

\section*{Conflicts of interest}

The authors declare no conflicts of interest.

\section*{Data availability}

All the data and calculations supporting this study's findings are available from the corresponding author upon reasonable request. The codes to recreate these data are available in a Zenodo repository \cite{jauma_exploring_2023}.

\section{Appendix}

\subsection{Parallel tempering}
\label{sec:comp}

We use the parallel tempering algorithm\ \cite{hukushima_exchange_1996}, which allows us to transform thermal averages into time averages sampled from a Markov chain in thermal equilibrium. This algorithm improves simulated annealing\ \cite{marinari_simulated_1992} by avoiding low-temperature configurations from getting trapped in local minima, thus providing a faster way of sampling thermal equilibrium probabilities at low temperatures. To run the algorithm, we simulate $N_r$ different \textit{realizations} of the couplings $J_{ij}$ in Eq.\ \ref{eq:model}, which will allow us to perform statistical averages. Additionally, we create $N_T$ \textit{replicas} of the Markov chain for each realization to perform the parallel tempering step, each replica being at a different temperature. 

Each Markov chain evolves through a series of Monte Carlo sweeps (MCS), where each sweep consists of running through all the spins, flipping their value---or not---according to the Metropolis rule, and then performing a parallel tempering step. In the latter, two replicas at different temperatures might exchange their spin configurations with a probability that ensures the detailed balance condition is met. In all of our simulations, we have ensured that the Markov chains are in thermal equilibrium by allowing the system to evolve during $N_t$ MCS (appendices \ref{sec:thermal-eq} explains how we determine whether $N_t$ is long enough or not to guarantee thermal equilibrium). Then, we obtain time averages by sampling the observable $N_t$ times:
\begin{equation}
\braket{\mathcal{O}} = \frac{1}{Z} \sum_S \mathcal{O}_S e^{-\beta E_S} \approx \frac{1}{N_t} \sum_{i=1}^{N_t} \mathcal{O}_i,
\end{equation}
where each sample is taken after one full MCS, and hence, the simulations have the time equally divided between thermalization and time averaging.

To calculate the spin overlap $q$ Eq.\ \ref{eq:spin-overlap}, we need to simulate two independent Markov chains for each temperature. No tempering swaps are allowed in these Markov chains to keep them independent, but the MCS rules stated above are applied in each set. That is, we run the parallel tempering algorithm twice to have two independent Markov chains dynamic at each temperature. The computational cost for each architecture corresponds to performing $N_r\times 2N_T$ different Markov chains.

We have run two small-scale simulations to optimize the parallel tempering parameters (as the number of replicas or the distance in temperatures between them) that we will use in each simulation. The first simulation uses small thermalization times $N_t$ and a small number of replicas small $N_T$ in order to obtain a rough estimate of the statistical average of the heat capacity $\overline{C_V}=\overline{\braket{U^2}-\braket{U}^2}/T^2$, which allows us to define a good set of temperatures for the parallel tempering method\ \cite{sabo_constant_2008}. This set of temperatures ensures that there is a constant increase in entropy and a constant probability of acceptance of tempering swaps between adjacent replicas. We then perform a second small-scale simulation, still with a small number of replicas but longer thermalization times. This second simulation is used to estimate the number of Monte Carlo sweeps $N_t$ to reach thermal equilibrium. Finally, we perform a longer-scale simulation with the parameters estimated in these pre-simulations. The details of these simulations can be found in table \ref{table:sim-params}.

\begin{table*}[t]
    \centering
    \begin{tabular}{|c|c|c|c|c|c|c|c|c c c|c|c|c|c|c|c|c|c|c|}
 \cline{1-8} \cline{12-19}
Model & Size & $T_\text{min}$ & $T_\text{max}$ & $N_T$ & MCS & $N_r$ & $T_c$ & & & & Model & Size & $T_\text{min}$ & $T_\text{max}$ & $N_T$ & MCS & $N_R$ & $T_c$ \\
\cline{1-8}                                                                              \cline{12-19}
2-RRG  &   100    & 0.2 & 1.5 & 30 & 80000   & 8695  &0.642 $\pm$ 0.005 & & & &   4-SW   & 100      & 1.0 & 2.5 & 27   & 160000  & 12391 &1.396 $\pm$ 0.007 \\
       &   200    & 0.2 & 1.5 & 30 & 80000   & 6762  &0.652 $\pm$ 0.004 & & & &          & 200      & 1.0 & 2.5 & 27   & 320000  & 6062  &1.447 $\pm$ 0.008 \\
       &   400    & 0.2 & 1.5 & 30 & 640000  & 49714 &0.667 $\pm$ 0.001 & & & &          & 400      & 1.0 & 2.5 & 28   & 640000  & 4108  &1.496 $\pm$ 0.008 \\
       &   800    & 0.2 & 1.5 & 30 & 1280000 & 23639 &0.675 $\pm$ 0.001 & & & &          & 800      & 1.0 & 2.5 & 28   & 1280000 & 10386 &1.533 $\pm$ 0.003 \\
       &   1600   & 0.2 & 1.5 & 30 & 5120000 & 2076  &0.687 $\pm$ 0.004 & & & &          & 1600     & 1.0 & 2.5 & 28   & 2560000 & 3745  &1.533 $\pm$ 0.005 \\
       & $\infty$ &     &     &    &         &       &0.733 $\pm$ 0.020 & & & &          & $\infty$ &     &     &      &         &       &1.634 $\pm$ 0.026 \\
\cline{1-8}                                                                     \cline{12-19}
4-RRG  &   100    & 0.5 & 3.0 & 30 & 320000  & 7020  &1.313 $\pm$ 0.008 & & & &   6-SW   & 100      & 1.3 & 3.0 & 32   & 160000  & 9556  &1.855 $\pm$ 0.009 \\
       &   200    & 0.5 & 3.0 & 30 & 320000  & 7562  &1.343 $\pm$ 0.006 & & & &          & 200      & 1.3 & 3.0 & 32   & 320000  & 4306  &1.919 $\pm$ 0.011 \\
       &   400    & 0.5 & 3.0 & 30 & 640000  & 1819  &1.363 $\pm$ 0.009 & & & &          & 400      & 1.3 & 3.0 & 32   & 640000  & 2972  &1.971 $\pm$ 0.011 \\
       &   800    & 0.5 & 3.0 & 30 & 1280000 & 2002  &1.393 $\pm$ 0.009 & & & &          & 800      & 1.3 & 3.0 & 32   & 1280000 & 4655  &2.009 $\pm$ 0.007 \\
       &   1600   & 0.5 & 3.0 & 30 & 5120000 & 2922  &1.417 $\pm$ 0.004 & & & &          & 1600     & 1.3 & 3.0 & 33   & 2560000 & 1399  &2.037 $\pm$ 0.010 \\
       & $\infty$ &     &     &    &         &       &1.508 $\pm$ 0.040 & & & &          & $\infty$ &     &     &      &         &       &2.144 $\pm$ 0.056 \\
\cline{1-8}                                                                    \cline{12-19}
6-RRG  &   100    & 0.5 & 3.0 & 30 & 160000  & 10374 &1.767 $\pm$ 0.008 & & & & Chimera & 72       & 0.2 & 3.0 & 30   & 320000  & 5012  & 1.146 $\pm$ 0.013 \\
       &   200    & 0.5 & 3.0 & 30 & 320000  & 6899  &1.802 $\pm$ 0.008 & & & &         & 200      & 0.2 & 3.0 & 30   & 640000  & 1987  & 1.092 $\pm$ 0.020 \\
       &   400    & 0.5 & 3.0 & 30 & 640000  & 3197  &1.844 $\pm$ 0.009 & & & &         & 392      & 0.2 & 3.0 & 30   & 640000  & 11123 & 1.067 $\pm$ 0.007 \\
       &   800    & 0.5 & 3.0 & 30 & 1280000 & 6638  &1.880 $\pm$ 0.005 & & & &         & 800      & 0.2 & 3.0 & 30   & 2560000 & 2672  & 0.983 $\pm$ 0.022 \\
       &   1600   & 0.5 & 3.0 & 30 & 5120000 & 6424  &1.901 $\pm$ 0.004 & & & &         & 1568     & 0.2 & 3.0 & 30   & 5120000 & 2508  & 0.904 $\pm$ 0.013 \\
                                                                                \cline{12-19}
       & $\infty$ &     &     &    &         &       &1.984 $\pm$ 0.028 & & & &  Pegasus & 128      & 0.2 & 4.0 & 43   & 320000  & 1950 & 2.291 $\pm$ 0.032 \\
\cline{1-8}
2-SW   &   100    & 0.5 & 2.5 & 30 & 160000  & 10010 &0.742 $\pm$ 0.006 & & & &          & 256      & 0.2 & 4.0 & 43   & 640000  & 2070 & 2.371 $\pm$ 0.029 \\
       &   200    & 0.5 & 2.5 & 30 & 320000  & 7520  &0.763 $\pm$ 0.005 & & & &          & 448      & 0.2 & 4.0 & 43   & 1280000 & 1551 & 2.355 $\pm$ 0.032 \\
       &   400    & 0.5 & 2.5 & 30 & 640000  & 3587  &0.799 $\pm$ 0.006 & & & &          & 960      & 0.2 & 4.0 & 43   & 2560000 & 1652 & 2.203 $\pm$ 0.029 \\
       &   800    & 0.5 & 2.5 & 30 & 1280000 & 1805  &0.831 $\pm$ 0.006 & & & &          & 1664     & 0.2 & 3.0 & 43   & 5120000 & 2192 & 2.130 $\pm$ 0.024 \\
                                                                                \cline{12-19}
       &   1600   & 0.5 & 2.5 & 30 & 2560000 & 5741  &0.851 $\pm$ 0.003 & & & &  Zephyr  & 48       & 0.5 & 5.0 & 40   & 80000   & 15952& 2.509 $\pm$ 0.025 \\
       & $\infty$ &     &     &    &         &       &0.926 $\pm$ 0.028 & & & &          & 160      & 0.5 & 5.0 & 40   & 80000   & 9704 & 2.862 $\pm$ 0.016 \\
       &          &     &     &    &         &       &                  & & & &          & 336      & 0.5 & 5.0 & 40   & 1280000 & 10407& 2.879 $\pm$ 0.016 \\
       &          &     &     &    &         &       &                  & & & &          & 576      & 0.5 & 5.0 & 40   & 2560000 & 3057 & 2.810 $\pm$ 0.030 \\
       &          &     &     &    &         &       &                  & & & &          & 1248     & 0.5 & 5.0 & 40   & 5120000 & 2448 & 2.661 $\pm$ 0.030 \\
\cline{1-19}
    \end{tabular}
    \caption{Simulation parameters for the different graphs that we have considered, the random regular graphs with $k$ branching number ($k$-RRG) and the small world graphs with $k$ average branching number ($k$-SW). The different columns correspond respectively to the simulation parameters defined above in this section: the number of spins of the Ising model \eqref{eq:model}, the minimum and maximum temperatures simulated, the number of replicas for the parallel tempering algorithm, the total number of Monte Carlo Sweeps, the total number of realizations and the estimation of the critical temperature with error bars. Note that for the random-regular and small-world graphs, the row with system size $\infty$ corresponds to the extrapolation to the thermodynamic limit using the $N^{-1/3}$ scaling.}
    \label{table:sim-params}
\end{table*}

\subsection{Estimation of autocorrelation times}
\label{appendix:correlation}

Here we briefly describe the scheme proposed in \cite{ambegaokar_estimating_2010} to calculate autocorrelation times and errors in thermal averages. We start with the original series of samples $A_{i}^{(0)}=A_i$ and then proceed to create new "binned" series by averaging over two consecutive entries. This is iteratively done as follows:
\begin{equation}
A_{i}^{(n)}={1\over 2}\left(A_{2i-1}^{(n-1)}+A_{2i}^{(n-1)}\right)
\end{equation}
where $i$ ranges from $1$ to $M_{n}\equiv M/2^{n}$, with $M$ the total number of bins at 0-th iteration. That is, $A_i^{(n)}$ contains the averages of two adjacent values in $i$ in the previous time series $(n-1)$.

We estimate the errors for each binned series $\Delta_A^{(n)}$ assuming the samples are uncorrelated, so they can be computed as:
\begin{equation}
\Delta_A^{(n)}
\approx
\sqrt{\frac{1}{M_{n}(M_n-1)}\sum_{i=1}^{M_{n}}\left(A^{(n)}_{i}-\overline{A^{(n)}}\right)^{2}}
\end{equation}

\begin{figure}[t]
    \centering
    \includegraphics[width=0.45\textwidth]{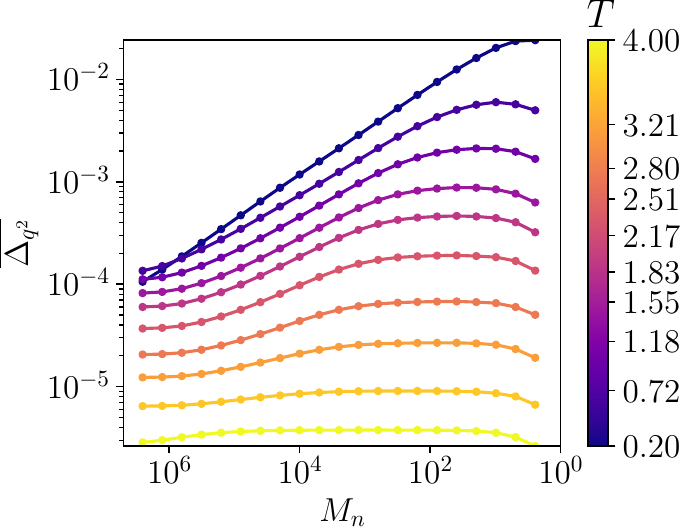}
    \caption{Thermal average error $\overline{\Delta_{q^2}}$ (right) as a function of the number of bins $M_n$ at the $n$-th iteration level of the procedure described in section \ref{appendix:correlation}. The convergence as $n$ increases for all temperatures below T=1.18 (note the plateaus) indicate that the system is thermalized. This example is for the Pegasus graph with $N=960$.}
    \label{fig:thermalization-test-3}
\end{figure}

The errors $\Delta_A^{(n)}$ increase as a function of the bin size $2^{n}$. However, they converge to the correct error estimate as the bins become uncorrelated for sizes $2^{n}\gg\tau_{A}$. The final error in the thermal average is thus given by the limiting value
\begin{equation}
\Delta_A = \lim_{n\rightarrow\infty}\Delta_A^{(n)};.
\end{equation}
However, note that in practice this formula breaks down for large $n$ when the number of bins $M_n$ is too small to calculate variances with statistical significance. This behavior can be observed in Fig. \ref{fig:thermalization-test-3} as a decrease in the estimation of the thermal error for the last iterations where the variances are calculated over a small number of bins.

Now that we have calculated the error in the thermal average of the variable under study, we can use it to calculate the autocorrelation time $\tau_{A}$ from the following relation:
\begin{equation}
\tau_{A}={1\over 2}\left[\left(\frac{\Delta_A}{\Delta_A^{(0)}}\right)^{2}-1\right].
\end{equation}

\subsection{Exponential fittings of autocorrelation times}
\label{appendix:exponential}
In the main text, we briefly mentioned the possibility that the correlation time $\tau_{q^2}$ could follow an exponential law \ref{sec:auto}. Such a possibility would introduce an additional degree of complexity in our analysis and may describe better our observations. However, the investigation of this law was limited by several factors, including the accuracy of autocorrelation time estimation and the lack of data for larger system sizes. In this appendix, we discuss further those issues.

\begin{figure*}[t!]
    \centering\includegraphics[width=\textwidth]{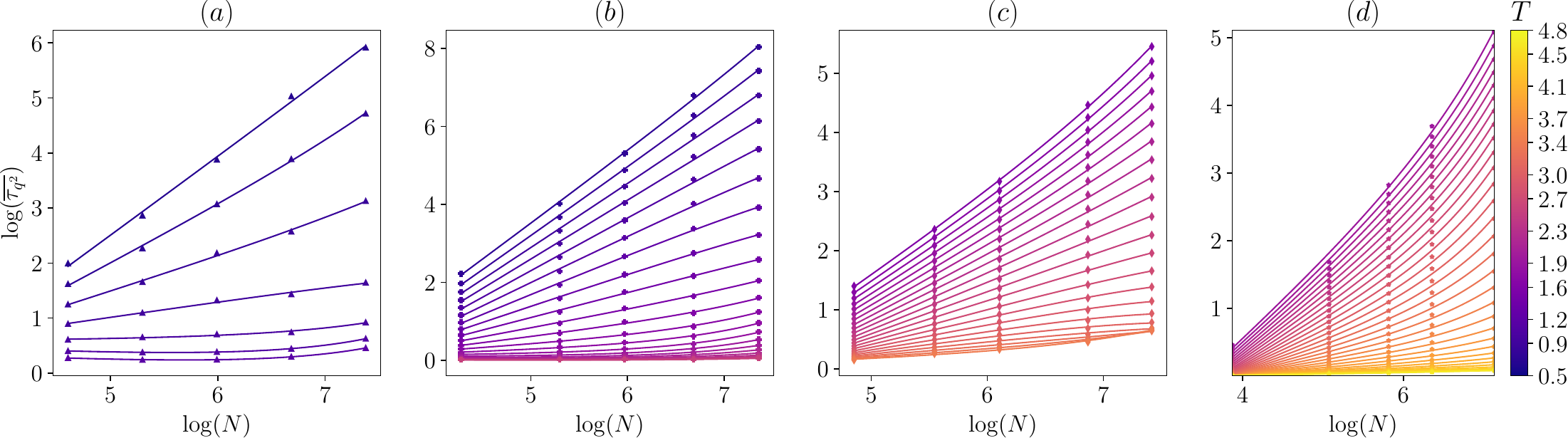}
    \caption{Log-log plot of autocorrelation time as a function of system size for different temperatures and graphs. The solid lines are a fit of the data points to the law $\tau_{q^2}=B(T)N^{A(T)}e^{C(T)}$.}
    \label{fig:autocorrelation-time-exp-law}
\end{figure*}

Figure \ref{fig:autocorrelation-time-exp-law} displays the log-log plot of the autocorrelation times as a function of system size for different temperatures and graphs. Each set of points corresponds to a specific temperature, and the lines represent fits to a law:
\begin{equation}
\tau_{q^2}=B(T)N^{A(T)}e^{C(T)}.
\end{equation}
In the log-log representation of Fig. \ref{fig:autocorrelation-time-exp-law}, a power-law relationship would correspond to a straight line ($C(T)=0$), while the curvature introduced by the exponential fit could potentially account for more complex dependencies between the correlation time and system size. As observed, the exponential fits seem to capture a change in the curvature when going from the paramagnetic to the spin-glass regime, suggesting that they might be more appropriate to describe the correlation time's behavior with system size. We also observe an additional tendency for the exponent $C(T)$ to become positive at large temperatures in Fig.\ \ref{fig:autocorrelation-time-exp-law}. This is probably caused by the fact that in this regime autocorrelation times are very short and we would need further system sizes to adequately characterize the low-temperature region of the phase diagram.

In any case, we cannot extract a solid conclusion about whether there is an exponential time divergence on the autocorrelation times from the data in Fig.\ \ref{fig:autocorrelation-time-exp-law} because there are large fluctuations in the resulting fitting parameters. Those fluctuations are caused by the fact that our system sizes are not large enough. Indeed, we are performing a fitting to a law with $3$ parameters (some of them are non-linear) using only $5$ data points with a small distance in $log(N)$ parameter. Furthermore, the error bars for the autocorrelation time itself are poorly estimated, making it difficult to use the goodness of fit to check the validity of our model. For all of this, we have characterized the appearance of a temperature where autocorrelation times significantly increase  with an easier-to-fit polynomial law, as explained in the main body of the text.

\subsection{Thermal equilibrium}
\label{sec:thermal-eq}
We employ the methods described in \cite{fernandez_critical_2008} and \cite{katzgraber_monte_2001} to test whether the Markov chains are in thermal equilibrium or not. As a first test, we perform a logarithmic binning of the time series used to estimate thermal averages---dividing the whole set of $N_t$ samples in bins such that the $n$th bin contains the samples in the interval $(N_t/2^{n+1},\: N_t/2^n]$, and then calculating the observables of interest using the samples of only one bin
\begin{equation}
    \braket{\mathcal{O}}_n =  \frac{1}{N_t^{n+1}} \sum_{i=N_t/2^{n+1}}^{N_t/2^n} \mathcal{O}_i
    \label{eq:bin-average}
\end{equation}

\begin{figure}[t!]
    \centering
    \includegraphics[width=.45\textwidth]{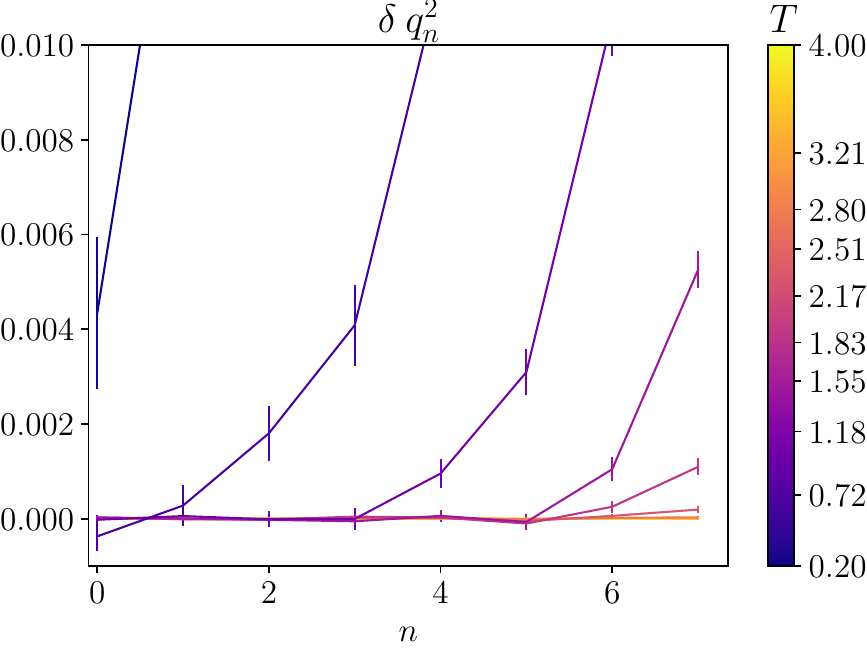}
    \caption{$\delta q^2_n = \overline{\braket{q^2}_0-\braket{q^2}_n}$ as a function of bin index (see eq. \ref{eq:bin-average}) for different temperatures of the Pegasus graph with $N=960$.}
    \label{fig:thermalization-test-1}
\end{figure}

If the Markov chain is in thermal equilibrium the observables must become $n$ independent for the first bins---the last and longest part of the simulation---and hence the statistical average of the difference between the observable of the first bin and the observables of subsequent bins, $\delta q^2_n = \overline{\braket{q^2}_0-\braket{q^2}_n}$, must converge to zero as $n$ is decreased, as shown in Fig. \ref{fig:thermalization-test-1}. Looking at this figure we can conclude that the minimum temperature at which this specific simulation has reached thermal equilibrium is for $T=1.18$ because $\delta q^2_n$ has converged to zero for the first 4 bins within error bars. This is the main method that we have used to check for thermalization, however, to verify our results we have also used several other tests.

\begin{figure}[t]
    \centering
\includegraphics[width=.45\textwidth]{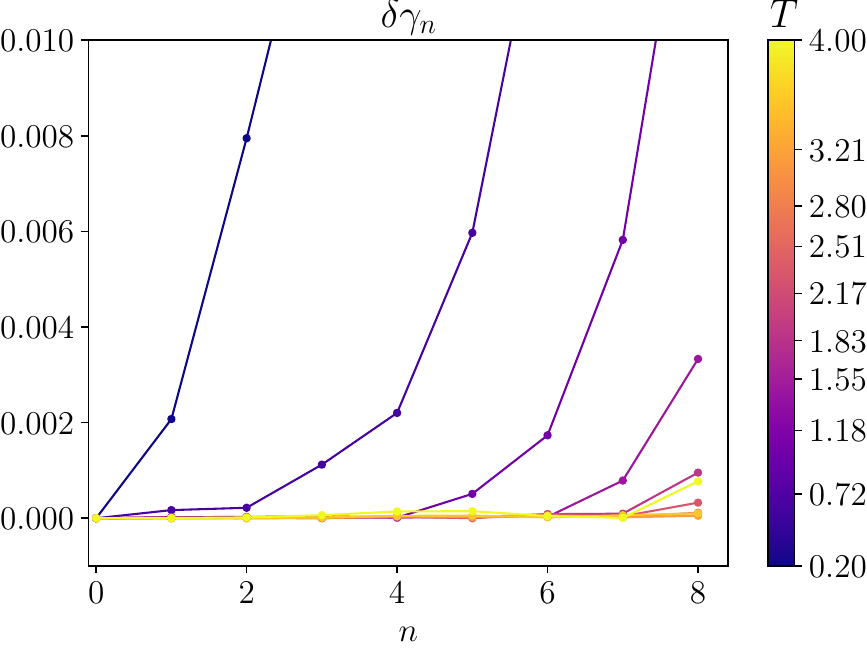}
    \caption{ $\delta \gamma_n = \gamma_0-\gamma_n$ as a function of bin index (see eq. \ref{eq:gamma}) for different temperatures of the Pegasus graph with $N=960$.}
    \label{fig:thermalization-test-2}
\end{figure}

A second method to test for thermal equilibrium is to repeat this test, but instead of studying a particular observable, we can check a relation between observables\ \cite{bray_observations_1980, katzgraber2001monte} that must be satisfied when the random couplings are obtained from a Gaussian distribution:
\begin{equation}
    \overline{\braket{q_l}} = 1-\frac{T|U|}{(z/2)J^2}
\end{equation}
where $J$ is the variance of the Gaussian distribution and $q_l$ is the link overlap, defined by 
\begin{equation}
    q_l = \frac{1}{N_b}\sum_{ \braket{ij} } S_i^\alpha S_j^\alpha S_i^\beta S_j^\beta \to \overline{\braket{q_l}}   =\frac{1}{N_b}\sum_{ \braket{ij} } \overline{\braket{S_i S_j}^2} \;,
\label{eq:link-overlap-relation}
\end{equation}
where $N_b = Nz/2$ is the total number of bonds between spins and $z$ is the coordination number---the average number of bonds per spin. Again, we can perform an analysis of this using the binding method and study the quantity
\begin{equation}
    \gamma_n = \overline{\braket{q_l}_n} -1+\frac{T|U|}{(z/2)J^2};,
    \label{eq:gamma}
\end{equation}
characterizing the convergence by comparing the result of the first bin and subsequent bins, $\delta \gamma_n = |\gamma_0-\gamma_n|$, 
where $\gamma_n$ corresponds to the evaluation of eq. \ref{eq:gamma} using the samples of the interval $(N_t/2^{n+1},\: N_t/2^n]$. According to Fig. \ref{fig:thermalization-test-2} we can again conclude that the minimum temperature for thermalization of the Pegasus graph with $N=960$ corresponds to $T=1.18$.

The binning method used to calculate thermal errors and autocorrelation times in the previous section can also be used as a third thermalization test. This is because in a thermalized Markov chain, the thermal errors should converge to a constant value as the size of the bins of the procedure explained in section \ref{appendix:correlation} increases. Our binning analysis provides a way to check for this convergence by plotting the estimated thermal errors as a function of bin size, as demonstrated in Fig. \ref{fig:thermalization-test-3}. The presence of a plateau in these plots, where the values cease to change significantly with increasing bin size, signifies that the Markov chain has reached thermal equilibrium. Specifically, we have adopted the criteria that the maximum of the thermal error must differ less than a $10\%$ from their adjacent points.

Furthermore, it's important to verify that the integrated autocorrelation time $\tau_A$ is significantly smaller than the total length of the time series $N_t$. This is to ensure that our Markov chain is "mixing" well and producing a sufficiently large number of effectively uncorrelated samples for reliable statistical analysis.

\begin{figure}[t]
    \centering
    \includegraphics[width=.45\textwidth]{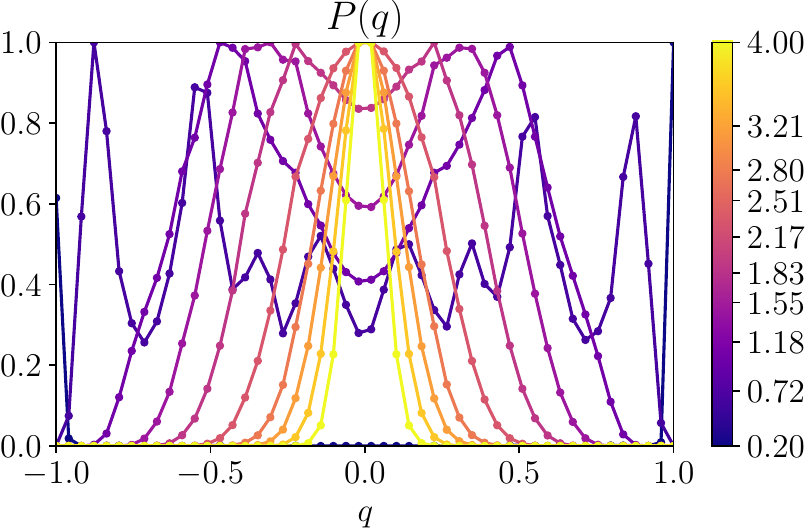}
    \caption{Probability distribution $P(q)$ of the spin overlap between two independent simulations at different temperatures, $q = \frac{1}{N}\sum_i S_i^\alpha S_i^\beta$. This example is for the Pegasus graph with $N=960$.}
    \label{fig:thermalization-test-4}
\end{figure}

The fourth and last thermalization test relies on the fact that we have imposed a symmetry in the coupling constants of the model, that is, $J_{ij}=J_{ji}$. Due to this, we know that in thermal equilibrium the distribution of the spin overlap $q$ must also be symmetric. Fig. \ref{fig:thermalization-test-4} shows $P(q)$ for different temperatures. We have not used this method to have an additional numerical thermalization test, but rather as a final check. Note that the distribution for $T=0.72$ is not completely symmetric, indicating (in agreement with our previous tests) that the system is not fully thermalized for that temperature.

% \pagebreak
\bibliographystyle{unsrt}
\bibliography{./references}

\end{document}